\documentstyle[11pt,newpasp,twoside,epsf]{article}
\def\edcomment#1{\iffalse\marginpar{\raggedright\sl#1\/}\else\relax\fi}
\reversemarginpar

\begin{document}
\title{Simulating a White Dwarf dominated Halo}
 \author{C. B. Brook, D. Kawata, B. K. Gibson}
\affil{Centre for Astrophysics and Supercomputing, Swinburne University of Technology, Hawthorn, VIC 3122, Australia}
\begin{abstract}
Halo initial mass functions (IMFs) heavily biased toward white dwarf (WD) precursors ($\sim 1-8$M$_{\odot}$) have been suggested as a suitable mechanism for explaining MACHO events. However, by simple chemical evolution argument, Gibson \& Mould (1997; GM97) pointed out that such WD-heavy IMFs (wdIMF) cause the overproduction of carbon and nitrogen. We re-examine this problem using numerical simulations.
\end{abstract}

\section{The Code \& the Model}
Numerical simulations are performed using an updated version of Kawata (2001). Metallicity dependant cooling is incorporated.  
Star formation occurs when $\rho>\rho_{crit}=2\times 10^{-25}gcm^{-3}$ \& $\triangle\cdot v_i<0$ and follows the Schmidt relation $SFR \propto \rho^{1.5}$. We use two IMFs, the canonical IMF of Salpeter (1955) (sIMF) and wdIMF of Chabrier (1996).  
Stars $> 10$M$_{\odot}$ explode as Type II supernova. We use metallicity dependent yields from Woosley \& Weaver (1995) for Type II SN ejecta, which are instantaeously fed back into the ISM. For stars $1-8$M$_{\odot}$ we use the yields of van den Hoek \& Groenwegen (1997) and recycling has a time delay of 1 Gyr to account for stellar lifetimes.

We calculate a semi-cosmological disk galaxy formation scenario after Katz \& Gunn (1991). Our seed galaxy is an isolated sphere upon which we superimpose small-scale density fluctuations corresponding to a CDM power spectrum. Longer wavelength fluctuations are incorporated by enhanced sphere density and initial solid-body rotation. We use 9171 gas and dark matter particles for a total galactic mass of $5\times 10^{11}$M$_{\odot}$ and examine the chemical evolution of our simulated galaxies using the sIMF, wdIMF and a variable IMF (vIMF). For the vIMF we use the wdIMF in low metallicity ($Z<0.05 Z_{\odot}$) star forming regions whilst we use the sIMF in high metallicity star forming regions.

\section{Results \& Conclusions}
Our simulations show that sub clumps of low metallicity stars become readily apparent by z=2, followed by a continued hierarchical collapse, while high metallicity stars are formed primarily at the time of disk formation. These large merger events occur in the epoch between z=1.1 and z=0.8. High metallicity stars preferentially end up in the disk and bulge regions, consistent with White \& Springel (2000).
Table 1 shows that the wdIMF scenario produced a factor of 5 more WDs than the sIMF model.  The number of WDs using the vIMF is similar to that of the wdIMF in the halo region and midway between the sIMF and wdIMF in the disk region. The vIMF yields a local halo WD mass fraction of $\sim 2\%$ consistent with the observations of Oppenheimer et al. (2001).

We examine [C/O] vs [Z] for our simulated stars. Observational constraints of [C/O] for halo stars ($[Z] < -1.5$) come from Timmes et al. (1995). Disk stars ($[Z]\sim 0$) have [C/O] $\sim 0$. Abundances of [C/O] for sIMF correspond to observations of both halo and disk stars. For wdIMF [C/O] increases and becomes too high once intermediate stars die and their high C yields are recycled. vIMF traces wdIMF in the low metallicity region. As metalllicity increases, carbon is not overproduced in vIMF nearly as dramatically as in wdIMF. [C/O] approaches that for sIMF as [Z] increases further, finally recovering [C/O]$\sim 0$ for disk stars.

The difficulty with reconciling a heavily WD populated halo with abundances of carbon is consistent with GM97. We find that the overabundance of carbon was mitigated by use of a simple variable IMF model in which sIMF was used in low metallicity star forming regions and wdIMF was used in high metallicity star forming regions. In addition our dynamical simulation reveals that vIMF leads to a high population of WDs in the halo in the process of hierarchical clustering.  

\begin{table}
  \caption{Number densities [$pc^{-3}$] of white dwarves in the 3 models. }
  \begin{tabular}{@{}cccc@{}}
    & vIMF & sIMF& wdIMF  \\
                   
 Halo density  &   $7.2 \times10^{-5}$ & $2.2  \times10^{-5}$        &    $ 9.8\times10^{-5}$ \\
Disk density  &   $ 2.7\times10^{-3}$ & $1.0\times10^{-3}$        &     $ 6.0\times10^{-3}$       \\
 
\end{tabular}
\end{table}

\end{document}